\documentstyle[12pt, onecolumn]{elsart}
\begin{document}


%

\def\simge{\mathrel{%
   \rlap{\raise 0.511ex \hbox{$>$}}{\lower 0.511ex \hbox{$\sim$}}}}
\def\simle{\mathrel{
   \rlap{\raise 0.511ex \hbox{$<$}}{\lower 0.511ex \hbox{$\sim$}}}}

\def\simeqq{\,\,{\raise 1.2ex\hbox{?}\llap{$\simeq$}}}

\def\+{\mathrel{
   \rlap{\raise -0.970ex \hbox{$\mathbf\widehat{}$}}
{\lower 0.511ex \hbox{$\mathbf\bigcirc$}}}}
\def\-{\,\,\,\,\,\mathrel{
   \llap{\raise -0.500ex \hbox{$\mathbf\bigcirc$}}
{\lower 0.990ex \hbox{$\mathbf\widehat{}$}}}}

\def\beq{\begin{eqnarray}}
\def\eeq{\end{eqnarray}}
\def\e{\mbox{e}}
%

\begin{frontmatter}

\title{\sc Robust flavor equalization of cosmic neutrino flux by 
quasi bi-maximal mixing \thanksref{C}}

\thanks[C]{Work supported by Consejo Nacional de Ciencia y 
Tecnolog\'ia (CONACyT) under Project \# 32067-E.}

\author{D.\ V.\ Ahluwalia, C.\ A.\ Ortiz, and G.\ Z.\ Adunas}

\address{Theoretical Physics Group, ISGBG, Ap. Pos. C-600\\
Escuela de F\'isica de la UAZ, 
Zacatecas, ZAC 98068, M\'exico.\\
E-mail: ahluwalia@phases.reduaz.mx; http://phases.reduaz.mx}

\date{23 June, 2000}

\maketitle


\begin{abstract}
For high energy cosmic neutrinos Athar, Je\'zabek, and Yasuda (AJY) 
have recently shown that the existing data on neutrino oscillations 
suggests that 
cosmic neutrino flux at the AGN/GRB source, 
$F(\nu_e):F(\nu_\mu):F(\nu_\tau)\approx 1:2:0$, oscillates to
$F(\nu_e):F(\nu_\mu):F(\nu_\tau)\approx 1:1:1$. These results can be
confirmed at AMANDA, Baikal, ANTARES and NESTOR, and other 
neutrino detectors with a good flavor resolution. Here, we re-derive the
AJY result from quasi bi-maximal mixing, and show that
observation of $F(\nu_e):F(\nu_\mu):F(\nu_\tau)\approx 1:1:1$ does not 
necessarily establish cosmic neutrino flux at the AGN/GRB source to be 
$F(\nu_e):F(\nu_\mu):F(\nu_\tau)\approx 1:2:0$. 
\end{abstract}


\end{frontmatter}


\newpage
{\it This preprint is no longer being pursued for publication.
Its essential contents  are now published 
as part of D. V. Ahluwalia, Ambiguity in source flux of cosmic/astrophysical 
neutrinos: Effects  of bi-maximal mixing and quantum-gravity 
induced decoherence,  Mod.\ Phys.\ Lett.\ A {\bf 16} (2001) 917
}

\section{Introduction}

The solar neutrino anomaly, the LSND excess events, and the Super-Kamiokande
data on atmospheric neutrinos, find their natural explanation in terms of 
oscillations of neutrinos from one flavor to another \cite{jb,LSND,SuperK}.
The only experiment so far that provides a direct evidence of oscillation 
from one flavor to another is the LSND experiment. However, the LSND 
result is still debated by the KARMEN collaboration \cite{k}. It is expected 
to be settled by the dedicated Fermi Lab. experiments. Nevertheless, a strong 
tentative evidence for neutrino oscillations seems  established.

In this {\em Letter\/} we shall neglect any possible CP violation in 
neutrino oscillations. We shall adopt the standard three-flavor neutrino 
scheme. In that framework one can accommodate any of the following two
sets of data: (a) Data on the atmospheric neutrinos
and solar neutrino anomaly, or (b)
Data on atmospheric neutrinos and LSND excess events. In the quasi bi-maximal
mixing, the angle $\theta$, see Eq.\ (\ref{bm}) below, can accommodate either
the LSND results or the solar anomaly, but not both.

In the context of this experimental setting, and the stated
theoretical framework, this {\em Letter\/} establishes the abstracted result. 
The origin for the abstracted result lies in the observation that
the observed $L/E$ flatness of the electron-like event ratio in the 
Super-Kamiokande atmospheric neutrino data strongly favors \cite{qbm1,qbm2} 
a quasi bi-maximal mixing matrix (and in fact this is what drives the AJY
result). Here we show that quasi
bi-maximal mixing transforms 
$F(\nu_e):F(\nu_\mu):F(\nu_\tau)\approx 1:a:2-a$ to
$F(\nu_e):F(\nu_\mu):F(\nu_\tau)\approx 1:1:1$.
Note, the latter flux neither carries an $a$ dependence, nor is
it affected by the angle $\theta$.
This robustness has the consequence that by studying the
departures from the 
$F(\nu_e):F(\nu_\mu):F(\nu_\tau)\approx 1:1:1$
for the observed cosmic 
high energy flux one may be able to explore new and interesting sources/physics
of high energy cosmic neutrinos. The data, however, may also be used to study
unitarity-preserving deformations of bi-maximality.  
 
In the next section,  we summarize the AJY result under study.
Section 3 shows the quasi bi-maximal mixing as the physical
origin of flux equalization for AGN/GRBs, it then presents the theorem
advertised in the {\em Abstract\/}, and it ends by introducing a deformed 
bi-maximal mixing and its affect on the flux equalization.
Section 4 is devoted to conclusion. 


\section{Brief review of AJY flux equalization} 

Without CP violation, the three-flavor neutrino oscillation 
framework carries five phenomenological parameters. These are the two
mass-squared differences, $\Delta m^{2}_{32}$ and $\Delta m^{2}_{21}$,
and the three mixing angles:

\def\ct{c_\theta}
\def\st{s_\theta}
\def\cb{c_\beta}
\def\sb{s_\beta}
\def\cp{c_\psi}
\def\sp{s_\psi}

\begin{equation}
U(\theta,\beta,\psi)=
\bordermatrix{
& 1 & 2 & 3 \cr
e   &\ct\cb & \st\cb & \sb \cr
\mu &-\ct\sb\sp - \st\cp & \ct\cp-\st\sb\sp & \cb\sp \cr
\tau    &    -\ct\sb\cp+\st\sp & -\st\sb\cp-\ct\sp & \cb\cp}
\end{equation}
The columns of the mixing matrix $U$ are numbered by the 
mass eigenstates, $\jmath=1,2,3$, while the 
rows are enumerated by the flavors, $\ell=e,\mu,\tau$. Here,
we have introduced the usual abbreviations: $c_x=\cos(x)$, and $s_x=\sin(x)$.

For a phenomenological study, the essential question is what are 
the parameters of the neutrino oscillations and what information 
may be extracted from them about particle physics, and astrophysical 
and cosmological processes/sources. New flavor-sensitive 
detectors with a collection area exceeding $1$ km$^2$ shall provide us
valuable information about the high-energy cosmic neutrino flux. This flux
carries important information about the conventional processes
of AGNs and GRBs, but it may  also serve as a probe of certain quantum
gravity effects and explore possible
violations of the equivalence 
principle\cite{qg1,qg2,qg3,qg4,qg5,qg6,qg7,qg8}.    

For high energy neutrinos, $E \simge 10^6$ GeV, with sources in AGNs
and GRBs, the source-detector distance far exceeds the kinematically
induced oscillation lengths suggested by any of 
the solar, atmospheric, and the LSND data. Under these circumstances
the AGN and GRB neutrino flux, $F^S$, at the source is roughly in the ratio:
\beq
F^S_e:F^S_\mu:F^S_\tau :: 1:2:0
\label{agn_grb}
\eeq
The oscillated flux, $F_\ell^D$, measured at terrestrial detectors,
becomes independent of the mass squared differences,
and is given by \cite{yasuda}:
\beq
F^D_\ell = \sum_{\ell^\prime}
P_{\ell\ell^\prime} F^S_{\ell^\prime}\label{yasuda}
\eeq
with
\beq
P_{\ell\ell^\prime} = \sum_\jmath \vert U_{\ell \jmath}\vert^2
 \vert U_{\ell^\prime \jmath}\vert^2
\label{p}
\eeq
Using the solar, reactor, atmospheric, and the accelerator, 
neutrino data AJY have made
a detailed numerical analysis. The result is \cite{yasuda}:\footnote{Also see, 
Ref. \cite{ajy}}
\beq
\mbox{AJY's numerical analysis:}\qquad
F^D_e:F^D_\mu:F^D_\tau :: 1:1:1
\label{num}
\eeq
Analytically \cite{ajy}, AJY show 
that the above result follows from the data-dictated assumptions:
\beq
\vert U_{e3}\vert^2 &\ll& 1, \nonumber\\
\left\vert \vert U_{\mu\jmath}\vert^2 -
\vert U_{\tau\jmath}\vert^2\right\vert &\ll& 1, \quad \jmath=1,2,3.
\label{ajy_anal}
\eeq 


\section{Quasi Bi-maximal origin of flux equalization and an 
ambiguity theorem}

We now show that this result is in fact a direct consequence
of the quasi bi-maximal mixing inferred from the $L/E$-flatness of the 
electron-like event ratio observed in the Super-Kamiokande data on
atmospheric neutrinos. Then, in the next section, we show
that the flux equalization is not a unique signature of the source flux
given by Eq.\ (\ref{agn_grb}).

It was argued in Refs. \cite{qbm1,qbm2} that the observed 
 $L/E$-flatness of the 
electron-like event ratio in the Super-Kamiokande data on
atmospheric neutrinos places severe {\em analytical\/} constraints on the
mixing matrix. Without reference to the solar neutrino deficit, or the
data on the LSND excess events, it was shown that these constraints
yield a quasi bi-maximal  mixing matrix.\footnote{The quasi 
bi-maximal mixing reduces to the bi-maximal mixing for $\theta=\pi/4$.
Apart from
Refs. \cite{qbm1,qbm2}, other
early references on bi-maximal mixing are \cite{bm1,bm2,bm3}.}  
This result is contained in Eq.\ (26) of Ref. \cite{qbm2}, and reads: 

\beq
U=
\left(\matrix{\ct & \st & 0 \cr
         -\st/\sqrt{2} & \ct/\sqrt{2} & 1/\sqrt{2} \cr
         \st/\sqrt{2} & -\ct/\sqrt{2} & 1/\sqrt{2} \cr}
\right)\label{bm}
\eeq
The mixing matrix (\ref{bm}), when coupled with Eq.\ (\ref{p}),  
yields:\footnote{An invertible quasi bi-maximal mixing matrix $U$, 
Eq.\ (\ref{bm}), necessarily yields a $P$ matrix that is non-invertible. 
This mathematical observation shall underlie the physical content
of the theorem to be presented below.}
\beq
P=
\left(
\matrix{\st^4+\ct^4 & \ct^2\st^2 & \ct^2\st^2 \cr
\ct^2\st^2 & \frac{1}{4}\left[1+\st^4+\ct^4\right] &
\frac{1}{4}\left[1+\st^4+\ct^4\right] \cr 
\ct^2\st^2   & \frac{1}{4}\left[1+\st^4+\ct^4\right] &
\frac{1}{4}\left[1+\st^4+\ct^4\right]}
\right)
\eeq
Substituting the obtained P in Eq.\ (\ref{yasuda}) furnishes with the 
prediction:
\beq
\mbox{Quasi Bi-maximal mixing:}\qquad F^D_e:F^D_\mu:F^D_\tau :: 1:1:1
\label{analytical}
\eeq
This is precisely the result (\ref{num}) which AJY obtained
based on a detailed numerical analysis \cite{yasuda}. On the analytical 
side \cite{ajy}, the AJY constraints (\ref{ajy_anal}) are manifestly
satisfied by the quasi bi-maximal mixing matrix (\ref{bm}).

Clearly, the AGN/GRB related ${\cal F}^S$ satisfy this flux equalization 
criterion with $a=2$. For supernovae explosions, $a\approx 1$. Once again,
one obtains the flux equalization. The early results of Learned and Pakvasa
\cite{lp}, and Weiler {\em et al.\/} \cite{tw}, are seen to follow as a special
case associated with $\theta=0$ and $a=2$.

The result (\ref{analytical}) is independent of the mixing angle
$\theta$ -- the angle relevant for the solar, or LSND, data (see  Refs. 
\cite{qbm1,qbm2}).
This implies that the high
energy cosmic neutrino flux is robust in that it does not depend
on the (vacuum) mixing angle obtained from the solar neutrino anomaly,
or from the LSND data. 
This robustness can be exploited to systematically study
other possible
significant sources of neutrino flux, especially those which may arise from 
sources other than the decay of charged pions. The latter component 
of the neutrino flux may appear as a departure from the 
evenly proportioned flux of the three neutrino flavors discussed here. 
The departures may also serve as a probe of certain quantum
gravity effects and possible
violations of the equivalence 
principle\cite{qg1,qg2,qg3,qg4,qg5,qg6,qg7,qg8}.    
However, we now emphasize that detecting a flux (\ref{analytical}) does not
necessarily imply the source flux to be (\ref{agn_grb}). 

In interpreting any deviations from the result (\ref{analytical}), one must 
be careful to note the following ambiguity theorem. Let 
\beq
{\cal F}^S \equiv F^S_e: F^S_\mu: F^S_\tau :: 1:a:2-a, \quad 0\le a
\le 2
\label{s}
\eeq
Then, under the already stated framework, the quasi bi-maximal mixing 
has the effect
\beq
{\cal F}^S \rightarrow {\cal F}^D
\eeq
where
\beq
{\cal F}^D \equiv F^D_e: F^D_\mu: F^D_\tau :: 1:1:1
\label{d}
\eeq
The proof is straight forward.

From an aesthetic point of view, a view which is also consistent
with the existing data, the quasi bi-maximal mixing is 
a strong candidate to emerge as the unitary matrix behind the neutrino 
oscillations. The widely discussed 
bi-maximal mixing \cite{qbm1,qbm2,bm1,bm2,bm3,wd}, as already
noted, is a special case of the quasi bi-maximal mixing. In this special case
one may introduce a unitarity-preserving deformation of the bi-maximality, 
and constrain it by the existing data as follows:   
\beq
U^\prime=
\left(\matrix{\cb/\sqrt{2} & \cb/\sqrt{2} & \sb \cr
               -(1+\sb)/2 & (1-\sb)/2 & \cb/\sqrt{2} \cr
               (1-\sb)/2 & -(1+\sb)/2 & \cb/\sqrt{2}}\right),\quad \beta\ll 1
\eeq
This deformed bi-maximal mixing transforms ${\cal F}^S$ given by 
Eq. (\ref{s}) into
\beq
{\cal F'}^D \equiv {F^\prime}^D_e: {F^\prime}^D_\mu: {F^\prime}^D_\tau :: 
1:1+(a-1)\sb^2:1+(1-a)\sb^2
\label{dd}
\eeq
and carries an essentially unique signature for the deformation parameter
$\beta$, and for the source flux parameter $a$ (associated with the class of 
neutrino fluxes under consideration).


\section{Conclusion}

The observed $L/E$ flatness of the electron-like event ratio in the 
Super-Kamiokande atmospheric data strongly favors a quasi bi-maximal mixing
for neutrino oscillations. This quasi bi-maximal mixing contains one 
unconstrained mixing angle, $\theta$. The angle $\theta$ can either 
be used to accommodate the LSND excess events, or to explain to the 
long-standing solar neutrino anomaly. For high energy cosmic neutrinos,
the Source-Detector distance far exceeds any of the relevant kinematically
induced oscillations lengths. When this information is coupled with the 
Super-Kamiokande implied quasi bi-maximal mixing, characterized by the angle
$\theta$, we find that a whole range of neutrino fluxes, 
${\cal F}^S$, defined in Eq.\ (\ref{s}), and characterized by $a$, oscillate to
equal fluxes of 
$\nu_e$, $\nu_\mu$, and $\nu_\tau$. This result carries a remarkable
robustness in its $\theta$- and $a$- independence. 

Observation of equal $\nu_e$, $\nu_\mu$, and $\nu_\tau$ fluxes from
AGN/GRBs, and supernovae explosions, can be used to establish if they 
belong to flux class, ${\cal F}^S$, defined above. Deviations
of these  fluxes,  ${\cal F}^D$, as observed in terrestrial detectors, 
from the ratio $1:1:1$ can become a robust measure of 
the departure
of the source flux ratio from $1:a:2-a$. A detailed study of these 
departures carries the seeds to discover new physics, and to characterize
cosmic neutrino sources. In particular, it is to be emphasized that $a$
remains unmeasurable, if the mixing is quasi bi-maximal (of which,
bi-maximal mixing is a special case). Furthermore,
the angle $\theta$, that, e.g., can be adjusted to resolve the solar 
neutrino anomaly,
does not influence the expected flux equalization. Because the high-energy
cosmic neutrino flux as detected in terrestrial laboratories 
is insensitive to the underlying mass-squared 
differences, measurements on the flavor spectrum
of the high-energy cosmic neutrino flux can be used to probe a 
whole range of  parameters associated with neutrino oscillations. Since each
of these parameters -- from those related to the 
deformed bi-maximal mixing,  to those that characterize a whole range of 
quantum-gravity effects (including those violating the principle of 
equivalence) -- is likely to yield a different
signature, high-energy cosmic neutrinos provide a powerful
probe into new physics.


\section*{Acknowledgements} 

One of us (DVA) extends his thanks to Sandip Pakvasa and Osamu Yasuda 
for remarks on the first draft of 
this {\em Letter\/} and for bringing to our attention some of the
early works on flux equalization in bi-maximal mixing.


\end{document}